\documentstyle[12pt]{article}
\newcommand{\be}{\begin{equation}}
\newcommand{\ee}{\end{equation}}

\newcommand{\ben}{\begin{eqnarray}\displaystyle}
\newcommand{\een}{\end{eqnarray}}
\newcommand{\refb}[1]{(\ref{#1})}

\begin{document}

{}~ \hfill\vbox{\hbox{hep-th/9602010}\hbox{MRI-PHY/07/96}
}\break

\vskip 3.5cm

\centerline{\large \bf $M$-Theory on $(K3\times S^1)/Z_2$}

\vspace*{6.0ex}

\centerline{\large \rm Ashoke Sen\footnote{On leave of absence from Tata
Institute of Fundamental Research, Homi Bhabha Road, Bombay 400005, INDIA}
\footnote{E-mail: sen@mri.ernet.in, sen@theory.tifr.res.in}}

\vspace*{1.5ex}

\centerline{\large \it Mehta Research Institute of Mathematics}
 \centerline{\large \it and Mathematical Physics}

\centerline{\large \it 10 Kasturba Gandhi Marg, Allahabad 211002, INDIA}

\vspace*{4.5ex}

\centerline {\bf Abstract}

We analyze $M$-theory compactified on $(K3\times S^1)/Z_2$ where
the $Z_2$ changes the sign of the three form gauge field,
acts on $S^1$ as a parity transformation and on $K3$ as an 
involution with eight fixed points preserving SU(2) holonomy. 
At a generic point in the moduli space the resulting theory has as its
low energy limit $N=1$ supergravity theory in six dimensions with eight 
vector, nine tensor and twenty hypermultiplets. The gauge symmetry
can be enhanced ({\it e.g.} to $E_8$) at special points in the moduli 
space. At other special points in the moduli space tensionless strings
appear in the theory.

\vfill \eject

\noindent {\bf Introduction:}
During the past year it has been realized that the moduli space of
string theories has a special point where the low energy dynamics
is governed by the $N=1$ supergravity theory in eleven 
dimensions\cite{TOWNSEND,WIS}. This theory has subsequently been called 
the $M$-theory, and it has been shown that some of the hidden symmetries
of string theory become apparent when we view it as
$M$-theory compactification on tori\cite{SCHWARZ,ASPIN}. 
More general compactifications of this theory on orbifolds of 
$S^1$\cite{HORWIT} and of more general tori\cite{MUKHI,WITTK3} have also
been studied and have been shown to be dual to known string 
compactifications. 

One of the main problems in studying $M$-theory on orbifolds is that
{\it a priori} there is no well defined rule for computing the
spectrum of twisted states, and one always has to rely on the 
requirement of anomaly cancellation to determine the massless
spectrum from the twisted sector. For theories with high number of
supersymmetries the spectrum of massless states is fixed more or
less uniquely by the requirement of supersymmetry and anomaly
cancellation. This feature has been exploited to determine the
spectrum of massless states in $M$-theory from the twisted sector.

In this paper we shall focus on $M$-theory on a $Z_2$ orbifold of
$K3\times S^1$. The $Z_2$ changes the sign of the three form field
$C_{MNP}$, acts as a parity transformation on $S^1$,
and acts as an involution on $K3$, $-$ the same involution used in
ref.\cite{SCHSEN} to construct the dual of CHL strings\cite{CHL}.
Even though the spectrum of massless states in this theory is not
completely fixed by the requirement of anomaly cancellation, we
shall be able to derive the spectrum by comparison with known
results about $M$-theory compactification on $T^5/Z_2$. The result
will be an $N=1$ supergravity theory in six dimensions with 
nine tensor multiplets, eight vector multiplets and twenty 
hypermultiplets, $-$ a theory for which the anomalies cancel
automatically\cite{SCHWARZAN,DABAN}.

\noindent{\bf The Model:}
Let us denote by $\tau$ the part of the $Z_2$ action that changes
the sign of $C_{MNP}$ and acts as a parity transformation on $S^1$,
and by $\sigma$ the involution on $K3$ that forms part
of the $Z_2$ action. Thus the $Z_2$ symmetry is given by the product
of $\tau$ and $\sigma$.  The involution $\sigma$ preserves $SU(2)$
holonomy and 
has eight fixed points. Its action on the lattice of signature (3,19),
representing the second
cohomology elements of $K3$, exchanges the two $E_8$ factors in the
lattice, leaving the (3,3) part invariant\cite{NIKULIN}.
Thus it has eight negative and fourteen positive
eigenvalues\cite{SCHSEN}. In particular it leaves the harmonic 
(0,2) and (2,0)
forms, as well as twelve of the harmonic (1,1) forms invariant, and
changes the sign of the other eight harmonic (1,1) forms. An example
of such a $K3$ surface is given by the hypersurface\cite{SCHSEN,CHLO}
\be \label{e1}
\sum_{i=1}^4 (z_i)^4 = 0\, 
\ee
in $CP^3$, where $z_i$ denote the homogeneous coordinates on
$CP^3$. The involution $\sigma$ is given by
\be \label{e2}
z_1\to - z_1, \qquad z_2 \to -z_2, \qquad z_3\to z_3, \qquad
z_4\to z_4.
\ee
The eight fixed points are at
\ben \label{e3}
z_1=z_2=0, \qquad (z_3/z_4)=\exp(2\pi i k/4), \qquad k\in Z\, ,
\nonumber \\
z_3=z_4=0, \qquad (z_1/z_2)=\exp(2\pi i k/4), \qquad k\in Z\, .
\een
Using Lefschetz fixed point theorem one can verify that this involution
leaves fourteen of the harmonic 2-forms fixed while changing the sign 
of the other eight\cite{SCHSEN}.

After we mod out by this $Z_2$ symmetry,
the resulting theory has $N=1$ supersymmetry in six dimensions. The
condition of anomaly cancellation by itself is not powerful enough
to determine the spectrum of massless states in the theory completely,
so we need to rely on some other principle for deriving the spectrum
of massless states in the twisted sector. The principle that we
shall adopt will be the following. Note that under the $Z_2$ transformation
the manifold $S^1\times K3$ has sixteen fixed points, $-$ two from $S^1$
combining with the eight from $K3$. Near each of these fixed points 
the space looks like $R^5/Z_2$ where the $Z_2$ changes the sign of all
the five coordinates. This is exactly the same structure that appears
at the orbifold points of $T^5/Z_2$. Thus we would expected that the
physics near these fixed points in $(K3\times S^1)/Z_2$ will be
identical to the physics near the fixed points of $T^5/Z_2$. The latter
case has already been analyzed in \cite{MUKHI,WITTK3}. In particular it
was shown in ref.\cite{WITTK3} that each of the fixed points acts as
a source of $-1/2$ unit of magnetic three form charge. The total 
magnetic charge is cancelled by putting $n_F/2$ five-branes moving on
the internal manifold where $n_F$ is the total number of fixed points.
Each of these five-branes in turn gives rise to a massless tensor
multiplet of the chiral $N=2$ supersymmetry algebra in six dimensions,
which corresponds to a tensor multiplet and a hypermultiplet of the
$N=1$ supersymmetry algebra. Since here $n_F=16$, we conclude that the
massless spectrum from the twisted sector corresponds to eight
tensor multiplets and eight hypermultiplets.\footnote{Note that
conventional heterotic string compactification on $K3$ only gives
theories with one tensor multiplet arising from the anti-symmetric
tensor field. But supergravity theories with more than one tensor
multiplet have been considered in the past\cite{AWA,ROMANS,SAG}.
The
possibility of getting extra tensor multiplets in six dimensional
theories by putting five-branes moving on internal manifolds has
already been discussed in ref.\cite{DUFF}. Here we have an explicit
realization of this idea. Other related six dimensional string 
compactification has been discussed in ref.\cite{GIM}.
} Following
\cite{WITTK3} we can also give a physical interpretation of the
scalars coming from the hypermultiplet and tensor multiplets. The
scalars in the tensor multiplet represent the location of the
five-branes along $S^1$, whereas those in the hypermultiplet
represent the location of the five-branes along $K3$.\footnote{Since
the $Z_2$ identification acts simultaneously on $S^1$ and on $K3$, we
see that the tensor- and the hyper- multiplet  moduli spaces do not
factorize globally.}

Having thus derived the spectrum of massless states in the twisted sector,
we now turn to the easier task of deriving the spectrum of massless
states in the untwisted sector. Let us denote by $\mu, \nu$ the 
six non-compact coordinates ($0\le \mu,\nu\le 5$), 
by $m,n$ the coordinates
on $K3$ ($6\le m,n\le 9$) and by 10 the coordinate on $S^1$. Then we
get one more tensor multiplet associated with the anti-selfdual component
of $C_{(10)\mu\nu}$ (the self-dual component being part of the gravity
multiplet), and eight vector multiplets associated with $C_{mn\mu}$
with $(mn)$ denoting one of the eight two cycles on $K3$ that are
odd under $\sigma$.
(Note that $C_{MNP}$ changes sign under the $Z_2$ involution). Finally
we get forty nine scalars (thirty four from the moduli of 
$\sigma$-invariant $K3$, one from the
radius of $S^1$ and fourteen from
$C_{(10)mn}$ with $(mn)$ denoting a two cycle of $K3$ even under
$\sigma$.) One of these scalars is part of the tensor multiplet
coming from the untwisted sector, the others form 12 hypermultiplets
of the $N=1$ supersymmetry algebra. The particular combination of
scalars which becomes part of the tensor multiplet coming from the
untwisted sector is given by
\be \label{enew}
\lambda = \sqrt{V\over R}\, ,
\ee
where $V$ is the volume of $K3$ and $R$ is the radius of $S^1$.

Combining the spectrum from the twisted and the untwisted sectors we
see that the massless sector corresponds to an $N=1$ supergravity 
theory in  six dimensions with nine tensor multiplets, eight vector
multiplets and twenty hypermultiplets. It can be easily seen that
this theory is anomaly free\cite{SCHWARZAN,DABAN}, which is a reflection
of the fact that $M$-theory has no gravitational anomaly on a
smooth manifold, and we have ensured that there is no anomaly from
the fixed points by putting $-1/2$ unit of magnetic charge at each
of the fixed points\cite{WITTK3}. Thus this provides another
example of a consistent $M$-theory compactification. In fact the
analysis based on $M$-theory provides us with a nice geometrical
picture of the hyper- and tensor- multiplet moduli spaces. The scalars
in twelve of the twenty hypermultiplets label the moduli space of
$\sigma$-invariant $K3$ together with background values of $C_{(10)mn}$
on cycles of $K3$ that are even under $\sigma$, those in the other eight 
hyper-multiplets 
label the location of the eight five-branes on $K3$, and the scalars
in the eight tensor multiplets coming from the twisted sector
label the locations of the eight five-branes on $S^1$. 

\noindent{\bf Enhanced Gauge Symmetries}:
Note that at a generic point in the moduli space there are no
massless charged states. There are however massive charged
states which arise from membranes wrapped around the two cycles
of $K3$. It is certainly possible
that they could become massless at special points in the moduli
space, but it would be difficult to predict such events directly, 
since the
spectrum of states in this theory is not controlled by Bogomol'nyi
bound. However, before the $Z_2$ projection, the theory had $N=2$
supersymmetry and hence BPS states; so we can first study the spectrum
in this theory and then project into $Z_2$ invariant states. As we
shall see, this provides a fruitful approach to determining points of
enhanced gauge symmetries in this theory.

In carrying out this analysis, we shall make use of
known dualities between the $M$ theory compactification on 
$K3\times S^1$ and heterotic compactification on $T^4$\cite{WIS}.
For our purpose it will be most convenient to make this duality
map in two stages; first regard $M$-theory on $S^1$ as a type IIA
theory, and then use the duality between type IIA theory on $K3$
and the heterotic string theory on $T^4$. In this language, the
transformation $\tau$ corresponds to the $(-1)^{F_L}$ symmetry of
the type IIA theory, which in turn, corresponds to a total inversion
of the $(4,20)$ lattice of charges in the heterotic string 
theory\cite{VAFAWIT}. On the other hand the transformation $\sigma$
acts by exchanging the two $E_8$ factors in the charge lattice of
the heterotic string theory\cite{SCHSEN}. Thus if we denote a vector
in the charge lattice of the heterotic string theory by
$(\vec k,\vec k_1, \vec k_2)$, where $\vec k$ is an eight 
dimensional vector representing momentum and winding in the 
compact direction, and $\vec k_1$ and $\vec k_2$ are vectors in the
two $E_8$ root lattices respectively, the net effect of the $Z_2$
action will be the map:
\be \label{e4}
(\vec k, \vec k_1, \vec k_2) \to 
(-\vec k, -\vec k_2, - \vec k_1)\, .
\ee 
Now consider being at a special point in the moduli space of heterotic
string theory on $T^4$ where
the $E_8\times E_8$ gauge symmetry is unbroken. In this case we
can get massless gauge bosons invariant under $Z_2$ by taking the
combination:
\be \label{e5}
|(\vec 0, \vec \alpha, \vec 0)\rangle_V + 
|(\vec 0, \vec 0, - \vec \alpha)\rangle_V\, ,
\ee
where $\vec \alpha$ is a root vector of $E_8$, and
the subscript $V$ denotes that we are considering space-time
vectors. 
We
also get $Z_2$ invariant massless charged scalars by considering 
the combination:
\be \label{e6}
|(\vec 0, \vec \alpha, \vec 0)\rangle_S - 
|(\vec 0, \vec 0, - \vec \alpha)\rangle_S\, .
\ee
The extra minus sign compensates for the fact that the vertex
operator for the scalar has a component proportional to one of the
right moving internal currents which are odd under $Z_2$. 
Similarly we can get massless charged spinors states.

Although we have constructed these states in  heterotic string
theory on $T^4$, by the chain of dualities, these states must
also exist in $M$-theory on $K3\times S^1$ at appropriate points
in the moduli space, and are $Z_2$
invariant there as well.\footnote{In carrying out the 
argument further, we are
implicitly assuming that in the theory obtained after the $Z_2$
modding, there is no phase transition between the
weak coupling region of $M$-theory, where the model was constructed,
and the weak coupling region of the heterotic string theory, where
one can see the appearance of massless charged states. To get further
insight, we note that the heterotic coupling constant $\lambda$
in six dimensions
is related to the volume $V$ of $K3$ and the radius $R$ of $S^1$
measured in the $M$-theory metric as $\lambda=\sqrt{V/R}$.  
Thus we can keep $\lambda$ small even when $R$ and $V$ are both large,
by keeping the ratio $V/R$ small. This of course does not mean that
the heterotic string theory is weakly coupled in this region, since one
or more of the radii of $T^4$ becomes large in this limit, so that
the correct description in the heterotic theory is as a higher
dimensional theory where it is strongly coupled. However, $\sqrt{V/R}$
becomes the scalar component of the tensor multiplet from the untwisted
sector after the $Z_2$ projection, and the moduli associated with
$T^4$ in the heterotic description become part of the hyper-multiplet
moduli space
after the projection. So the above analysis does show that
we can go from the region of weakly coupled $M$-theory to 
the region of weakly
coupled heterotic theory by remaining at a fixed point in the
tensor-multiplet moduli space ({\it i.e.} keeping $V/R$ fixed), 
and moving in the hyper-multiplet moduli
space. Thus we do not expect phase transitions of the kind 
discussed in ref.\cite{DUFF} to appear as we move from the region
where the $M$ theory is
weakly coupled to the region where the heterotic string theory is
weakly coupled.}
Thus when we make the $Z_2$ projection, 
these states are expected to survive, and will give us massless states
in the resulting theory. (As we shall argue later, quantum corrections
do not make these states massive.) In particular the  
massless charged vector states, together with appropriate fermionic
states, and the eight $U(1)$ vector
multiplets, give us the $E_8$ gauge multiplet. 
On the other hand, the massless charged scalars, together with
appropriate spinors, and eight neutral hypermultiplets associated
with $Z_2$ invariant $E_8\times E_8$ Wilson lines (in the
language of heterotic string theory), will
form a hypermultiplet in the adjoint representation of $E_8$.
Thus the resulting theory will have nine tensor multiplets,
an $E_8$ vector multiplet, a hypermultiplet in the adjoint
representation of $E_8$, and twelve neutral hypermultiplets as its 
massless spectrum. 

Are there new massless charged 
states coming from the twisted sector? To answer this
question, we need to look at the geometry of 
$K3$ near the enhanced symmetry
points. As has been argued in refs.\cite{WIS,STROM,BER}, in $M$-theory on 
$K3\times S^1$, enhanced gauge symmetries
occur at points in the $K3$ moduli space where some two cycles  collapse
to zero size. In particular, the enhanced $E_8\times E_8$ gauge symmetry
will occur at a point where two sets of non-intersecting two cycles, 
$-$ one associated
with the first $E_8$, and the other associated with the second $E_8$, $-$
collapse to zero size. The action of the involution
$\sigma$ does not leave any of these
two sets of cycles invariant, 
but exchanges them. As a result the fixed point
singularities introduced due to
the orbifolding by $Z_2$
are distinct from the singularities in $K3$ arising from collapsed
two cycles. 
Thus we would expect that the massless spectrum from the 
twisted sector, associated with the fixed points of $\sigma$, will not
be affected by the collapse of the two cycles. 
This argument continues to hold even when we move the
five-branes away from the fixed points, at least as long as they
do not hit the collapsed two cycles.
The net result of this is that the twisted sector does not 
generically give rise to new charged massless states in the spectrum. This
is just as well, since the spectrum of states that have already been found,
gives an anomaly free theory.

Since in this case there is no supersymmetry to protect the massless
charged states against gaining mass (by absorbing the charged
hypermultiplets), we must examine carefully the fate of these
states under quantum corrections. Although quantum corrections in
$M$-theory are not properly understood yet, we can make the
following observation. Let us consider the case where both $K3$
and $S^1$ have large volume, but we are at the $(E_8\times E_8)$
singular point of $K3$. In this case, before the $Z_2$ projection
the massless states come from membranes wrapped around the 
vanishing two cycles of $K3$. Thus these states are localized on
$K3$, but not on $S^1$, and we obtain a set of
states in the six dimensional theory by quantizing propagation
of these massless states on the internal
$S^1$. Before the $Z_2$ projection,
there is enough supersymmetry to guarantee that the zero modes
of these fields along $S^1$ give rise to zero mass states in the
resulting six dimensional theory (even after taking into account
interactions). Now let us consider the effect
of $Z_2$ modding. As has been argued before, $Z_2$ has the effect
of exchanging two sets of vanishing cycles on $K3$. Thus if we
denote by $M$ the neighbourhood of the singular point on $K3$,
then after the $Z_2$ modding,
the internal space still has  the structure $M\times S^1$ near the
singular point on $K3$. As a result,
the spectrum of the resulting six dimensional theory is still
obtained by quantizing propagation of the massless charged fields
on the internal $S^1$.
Since this is the same problem as before, we would expect that the
zero modes of these fields on $S^1$ would still produce massless
states in the resulting six dimensional theory. Put another way,
the quantum problem that we need to solve to determine if the
spectrum contains the massless charged states do not know about
supersymmetry breaking, and hence will continue to give
massless charged states in the theory even after the $Z_2$ 
modding.\footnote{In principle the world volume
instantons\cite{BBK}, wrapped around $S^1$, and a two cycle on $K3$ that
encloses both sets of collapsed two cycles, can give an exponentially
small contribution to the mass of these charged particles. This is
possible since such configurations do `know' about the $Z_2$
projection. In
this case we would have to interprete this as the charged
vector multiplet (and the 
charged hypermultiplet) acquiring small mass through non-perturbative
effects in $M$-theory. Put another way, the hypermultiplet moduli
space will contain a point where the gauge symmetry is almost restored
(in the sense that the charged vector- and hyper- 
multiplets have small mass) but is never completely restored.
E. Witten has suggested that it might be
possible to argue that such a phenomenon does not occur in this
theory, since it will be hard to write down a low energy effective
field theory satisfying this requirement.}
Of course, in this case we shall get only half of the states
compared to what we had previously, since instead of having two
copies of the singular region $M\times S^1$, we now have only one
copy.

The point of enhanced gauge symmetry also gives us a way of
determining the coupling between the tensor and vector
multiplets. For this let us consider a limit where we increase the size
of $K3$ by remaining at the point of enhanced $E_8\times E_8$
symmetry, and at the same time, adjust the radius $R$ of $S^1$ so
as to keep the scalar in the tensor multiplet, coming from the
untwisted sector, fixed. In this case the coupling between the
states in the twisted sector
tensor multiplet, which live on the five-branes,
and the charged vector multiplets, which live on the collapsed
two cycles of $K3$, can be made
to vanish by taking the location of the five branes on $K3$ to be
far away from the location of the collapsed two cycles on $K3$.
But this coupling cannot depend either
on the location of the five-brane on $K3$, or the size of $K3$ 
(appropriately scaled by the radius of $S^1$),
since these belong to the hypermultiplet moduli space. Thus the coupling
between the tensor multiplets coming from the twisted sector, and
the $E_8$ gauge multiplets (including the $U(1)^8$ factor) must
vanish identically. Only the tensor multiplet belonging to the 
untwisted sector couples to the vector multiplets. Using the fact
that the anomaly polynomial vanishes, and the results of
ref.\cite{SAG}, this coupling can be shown to be of the form
$\sqrt{-g} e^{-\Phi} tr(F^2)$, where $g$ denotes the canonical
metric, and $\Phi(=\ln\sqrt{V/R})$ is the scalar component of the
tensor multiplet in the untwisted sector.

One might try to get further enhancement of gauge groups by
going to
special points in the moduli space of the (4,4) lattice where the theory
before the $Z_2$ projection develops enhanced gauge symmetries.
If $\vec \beta$
denotes any non-zero element of such a special 
$(4,4)$ lattice representing a massless
charged particle in the theory before the $Z_2$ projection,
then we can
get new $Z_2$ invariant massless vectors by considering the
combination
\be \label{e7}
|(\vec \beta, \vec 0, \vec 0)\rangle_V + 
|(-\vec \beta, \vec 0, \vec 0)\rangle_V\, .
\ee
Similarly, $Z_2$ invariant massless scalars are obtained by taking
the combination
\be \label{e8}
|(\vec \beta, \vec 0, \vec 0)\rangle_S - 
|(-\vec \beta, \vec 0, \vec 0)\rangle_S\, .
\ee
The spectrum obtained this way is free from gravitational
anomalies since we get equal number of vector- and hyper- multiplets.
However, typically the spectrum suffers from gauge anomalies.
Consider for example, going to a special point in the $(4,4)$
lattice where the theory before the $Z_2$ projection develops
an enhanced $SU(n)$ gauge symmetry ($n\le 5$). 
The action of $Z_2$ on the group 
elements can be represented as complex conjugation of the $SU(n)$
matrices. Thus the gauge group after the $Z_2$ projection will
be $SO(n)$, generated by real $SU(n)$ matrices. The vectors
given in \refb{e7}, together with appropriate spinors, constitute
the vector multiplet in the adjoint of $SO(n)$. On the other
hand, the scalars given in \refb{e8}, together with the moduli
fields that need to be tuned to reach the $SU(n)$ point, and
appropriate spinors, constitute a hypermultiplet in the symmetric
rank two tensor representation of $SO(n)$.

Is there a possibility of getting massless states from the twisted
sector?
In the $M$-theory on $K3\times S^1$,
enhanced gauge symmetries of the type discussed above occur
at points in the $K3$ moduli space where a set of 2-cycles
associated with the $(3,3)$ part of the lattice of second
cohomology elements collapses to zero size.\footnote{This can
also involve special limit where the size of $K3$, measured in
type IIA metric, is of order unity, and the $C_{(10)mn}$ fields
are adjusted appropriately. This would correspond to enhanced
symmetries appearing from the (1,1) part of the lattice that is
not associated with the second cohomology elements of $K3$, but
to the zeroth and fourth cohomology elements.}
These cycles are mapped on to themselves under the action of
$\sigma$. As a result, the singularities introduced
by the orbifolding procedure are certainly going to interfere with
the singularities introduced on $K3$ by collapsing two cycles,
and we would, in general, expect new charged massless states from
the twisted sector. 

It is conceivable that the spectrum from the
twisted sector arranges itself so as to give an anomaly free theory. 
However, as we shall now argue, in this case there is no 
guarantee that the extra charged massless states obtained from
the untwisted sector survives quantum corrections, and hence a more
likely possibility is that we do not get any gauge symmetry
enhancement at all in the quantum theory. To see this we follow the
same approach as in the case of the $E_8\times E_8$ singular points.
In this case the massless charged states are again localised on
$K3$ near the vanishing cycles. Let us denote by $N$ the region
in $K3$ near the vanishing cycles. Thus before the $Z_2$ projection 
the internal space near the singular point had the structure
$N\times S^1$, and the 
quantum problem to be solved involved the propagation of the
massless charged fields on an internal
$S^1$. Supersymmetry guaranteed that the
zero modes of these massless fields along $S^1$ give rise to
strictly massless states in six dimensions. Situation after the
$Z_2$ modding is quite different however. Since in this
case the region $N$ is mapped onto itself under the $Z_2$ action,
the region near the singular point after the $Z_2$ modding
no longer has the structure $N\times S^1$. In particular, if
$P$ denotes the singular point onto which the two cycles have
collapsed, then at $P$ the internal space has the form
$P \times (S^1/Z_2)$. 
Thus in this case the quantum problem to be solved
for determining the spectrum of massless charged states
in the six dimensional theory
reduces to studying the propagation of these massless
fields not on an internal $S^1$, but on an internal $S^1/Z_2$.
This of course is a different problem, and now there is not
enough supersymmetry in the problem to guarantee that we get 
massless states in the six dimensional theory even after taking
into account quantum corrections. Thus we see that the massless
charged states that we have constructed in the untwisted sector, 
do not in general remain
massless after we take into account quantum corrections.

Interesting phenomena might occur at other points in the moduli
space where the heterotic string theory develops more general
enhanced gauge symmetries involving the full $(4,20)$ lattice.
But again in this case one or more of the collapsed cycles are
invariant under the $Z_2$ action, and hence our ability to
analyze the spectrum of massless states breaks down.

\noindent{\bf Duality Symmetries:} 
This theory has several $T$-duality symmetries. One class of such
symmetries involve the permutation of the eight five-branes, which
acts on the eight tensor multiplets arising from the twisted sector
as permutation group. We also have a set of duality symmetries that
exchange the eight vector multiplets among each other; $-$ this
is part of the global diffeomorphism group $O(3,19;Z)$ of $K3$
that commutes with the $Z_2$ projection.

Upon compactification on a circle, both the vector and the tensor
multiplets give rise to vector multiplets in the resulting five
dimensional theory. In the spirit of ref.\cite{HT}
it is tempting to conjecture that this theory has
a $U$-duality group that exchanges the vectors coming from the
six dimensional vector multiplet with the ones coming from the
six dimensional tensor multiplet. It will be interesting to
study what kind of moduli space\cite{SIER} arises as a result of
this compactification.

Finally one might ask if the $M$-theory compactification discussed
here has a string theory dual. It is clear that the dual in this
case cannot be a heterotic string compactification on $K3$ since
that always gives only one tensor multiplet. It turns out that
Dabholkar and Park have independently constructed six dimensional
string theories with precisely the same spectrum of massless
states at a generic point in the moduli space\cite{DABH}. 
Thus we expect these $M$-theory compactifications to be dual to the 
ones constructed in ref.\cite{DABH} in a non-perturbative sense.

\noindent {\bf Tensionless strings}: There are several apparently 
singular regions in the
moduli space of this theory which deserve further study. One of these,
already discussed earlier, corresponds to the region where the
$K3$ becomes singular due to collapsed two cycles, and the
locations of some of 
the fixed points of $K3$ under the involution $\sigma$ coincide
with the locations of the collapsed two cycles. Another interesting
region would be where the locations of two or more five-branes in
the internal space coincide. There can be open membranes
stretched between two five-branes in the $M$-theory\cite{STOP,TOOP},
which would appear as self-dual strings to a six dimensional
observer. These string are expected to become tensionless in the
limit where the relative separation of two or more five-branes
vanish\cite{STOP}, giving rise to a phenomenon similar to the one
observed in type IIB string theory compactified on $K3$\cite{WITJUL}.
Note that reaching these special points in the moduli space 
requires simultaneous adjustment of hyper-multiplet moduli (the
locations of the five-branes on $K3$) and tensor multiplet moduli
(the locations of the five-branes on $S^1$.) Upon further
compactification on a circle, these tensionless strings, wrapped
around the extra circle, will give rise to new massless states in
the theory, and might further enhance the gauge symmetry. This
lends further support to the conjecture that the five dimensional
theory might have a duality symmetry that exchanges the vectors
arising from the six dimensional vector multiplet with those
arising from the six dimensional tensor multiplet.

\noindent{\bf Acknowledgement:} I would like to thank A. Dabholkar,
J. Schwarz, A. Strominger and E. Witten
for useful discussions. Part of this work was carried
out at the Institute for Theoretical Physics at Santa Barbara, and was
supported in part by the National Science Foundation under grant No.
PHY94-07194.

\end{document}